\documentclass[twocolumn, preprint2]{aastex61}

\usepackage{graphicx}
\usepackage{amsmath}

\usepackage[normalem]{ulem}

\newcommand{\fb}[1]{{ #1}}
\newcommand{\stth}[1]{}

\newcommand{\RBSL}{{\footnotesize R}BS{\footnotesize L}}
\received{October 25, 2017}
\revised{December 4, 2017}
\accepted{???}

\submitjournal{ApJL}

\shorttitle{Regularized Biot-Savart Laws}
\shortauthors{Titov et al.}

\begin{document}

\title{Regularized Biot-Savart Laws for Modeling Magnetic Flux Ropes} 


\correspondingauthor{Viacheslav S. Titov}
\email{titovv@predsci.com}

\author{Viacheslav S. Titov}
\affiliation{Predictive Science Inc., 9990 Mesa Rim Road, Suite 170, San Diego, CA 92121} 

\author{Cooper Downs}
\affiliation{Predictive Science Inc., 9990 Mesa Rim Road, Suite 170, San Diego, CA 92121} 

\author{Zoran Miki\'{c}}
\affiliation{Predictive Science Inc., 9990 Mesa Rim Road, Suite 170, San Diego, CA 92121} 

\author{Tibor T\"{o}r\"{o}k}
\affiliation{Predictive Science Inc., 9990 Mesa Rim Road, Suite 170, San Diego, CA 92121} 

\author{ Jon A. Linker}
\affiliation{Predictive Science Inc., 9990 Mesa Rim Road, Suite 170, San Diego, CA 92121} 

\author{ Ronald M. Caplan}
\affiliation{Predictive Science Inc., 9990 Mesa Rim Road, Suite 170, San Diego, CA 92121}

\begin{abstract}
Many existing models assume that magnetic flux ropes play a key role in solar flares and coronal mass ejections (CMEs).  It is therefore important to develop efficient methods for constructing flux-rope configurations constrained by observed magnetic data and the morphology of the pre-eruptive source region.  For this purpose, we have derived and implemented a compact analytical form that represents the magnetic field of a thin flux rope with an axis of arbitrary shape and circular cross-sections. This form implies that the flux rope carries axial current $I$ and axial flux $F$, so that the respective magnetic field is the curl of the sum of axial and azimuthal vector potentials proportional to $I$ and $F$, respectively.  We expressed the vector potentials in terms of modified Biot-Savart laws whose kernels are regularized at the axis in such a way that, when the axis is straight, these laws define a cylindrical force-free flux rope with a parabolic profile for the axial current density.  For the cases we have studied so far, we determined the shape of the rope axis by following the polarity inversion line of the eruptions' source region, using observed magnetograms.  The height variation along the axis and other flux-rope parameters are estimated by means of potential field extrapolations.  Using this heuristic approach, we were able to construct pre-eruption configurations for the 2009 February 13 and 2011 October 1 CME events.  These applications demonstrate the flexibility and efficiency of our new method for energizing pre-eruptive configurations in simulations of CMEs.
\end{abstract}


\section{Introduction}
 \label{intro}

Coronal mass ejections (CMEs) are large eruptions of magnetized plasma from the solar corona into the heliosphere, and the main driver of geomagnetic storms.
There exists little doubt that CMEs consist of magnetic flux ropes \citep[FRs; e.g.,][]{Chen2017}, although the time of their formation has been debated.
In recent years,
however, there is growing evidence for the existence of FRs before many eruptions
\citep[e.g.,][]{Canou2009, Green2009, Zhang2012, Patsourakos2013, Howard2014, Chintzoglou2015}.
This justifies the use of FR configurations as initial condition for the numerical modeling of CMEs, for both idealized configurations \citep[e.g.,][]{Amari2000, Fan2005, Aulanier2010, Torok2011} and configurations that are constructed using observed magnetograms \citep[e.g.,][]{Manchester2008, Linker2016b}.

Modeling pre-eruptive FRs for observed cases is challenging, since direct measurements of the coronal magnetic field are difficult, so typically the morphology and magnetic structure of the FR can be inferred only indirectly from, e.g., observed filament shapes or the location of flare arcades or dimmings \citep[e.g.,][]{Palmerio2017}.
Multiple trial and error attempts may be required to create a stable magnetic equilibrium that satisfactorily matches the observations.
In contrast, a less rigorous approach based on out-of-equilibrium FRs for initializing CMEs is much easier to apply \citep[e.g.,][]{Liu2008, Lugaz2009, Loesch2011}.
However, there is no guarantee that the resulting CME model will be sufficiently accurate, especially for complex pre-eruptive configurations.

One way to produce equilibrium FR configurations is by mimicking the slow formation of pre-eruptive FRs using a boundary-driven MHD evolution \citep[e.g.,][]{Lionello2002, Bisi2010, Zuccarello2012, Jiang2016}.
This requires the development of photospheric boundary conditions that will lead to the formation of an FR, subject to the constraints of the observed photospheric field.
This approach is non-trivial, computationally expensive, and has no simple means to control the shape and stability of the FR.

Another way to produce such configurations is via non-linear force-free field \citep[NLFFF, e.g.,][]{Schrijver2008} extrapolations.
This method requires the use of observed vector magnetic data.
However, these data are measured at the photospheric level where the magnetic field is not force-free,
and must be ``pre-processed'' to be compatible with the force-free reconstruction \citep{Wiegelmann2006}.
In reality, the magnetic field becomes force-free in a thin layer where the photosphere turns into the chromosphere.
Vector magnetic field observations also suffer from noise and disambiguation issues, making constructing NLFFF-models of FRs that match observable properties a rather non-trivial problem as well.

An alternative approach is the FR insertion method \citep{vanBall2004,Su2011,Savcheva2012}, which uses observations of filaments, loops, etc., to directly constrain the field model.
For this technique, an attempt to find an equilibrium configuration is made in two steps.
First, a field-free cavity following the inferred FR shape is prepared in the related potential-field configuration and filled with axial and azimuthal magnetic fluxes.
Second, the resulting configuration is subjected to a magnetofrictional relaxation to find a force-free equilibrium.
The procedure is repeated by varying the inserted magnetic fluxes and/or cavity until an equilibrium with desirable properties is obtained.
Each initial configuration is highly out of equilibrium, since it is constructed without balancing the magnetic forces in the cavity.
As a result, the relaxation of the configuration can significantly change the inserted fluxes and perturb the ambient structure around the cavity. This makes the properties of the modeled force-free equilibrium difficult to control, and many parameter iterations may be required to reach the desired result.

In contrast, the equilibrium conditions are addressed in the FR embedding method by \citet{Titov2014}, which employs
\fb{a modified version of the model by \citet{Titov1999}, called henceforth TDm model.
}
This method uses the ambient potential field to estimate the axial and azimuthal magnetic fluxes, as well as geometric parameters of the FR.
The estimation \fb{follows from the requirement that 
the ambient field superimposed on the FR field must compensate the field component due to FR curvature, yielding an approximately force-free configuration}. 
This configuration is then relaxed via a line-tied MHD evolution toward an equilibrium whose parameters are close to the estimated ones---a main advantage of the method.

Unfortunately, since the TDm FR has a toroidal-arc shape, it is difficult to model configurations that reside above a highly elongated or curved polarity inversion line (PIL).
Complex FR shapes can be obtained by merging several FRs \citep[e.g.,][]{Linker2016b}, but this is a rather laborious procedure.
The purpose of the present work is to remove this geometric limitation by generalizing the method for FRs of arbitrary shape.


\section{Regularized Biot-Savart laws}

We propose a general mathematical form that defines at a given point ${\pmb x}$ the magnetic field ${\pmb B}_{\mathrm{FR}}$ of a thin FR with axis path $\cal C$, arc length $l$, radius-vector $ {\pmb R}(l)$, tangential unit vector ${\pmb R}^{\prime} ={\rm d} {\pmb R}/{\rm d}l$, and cross-sectional radius $a(l)$
(Figure~\ref{f:FR_3d}) as follows:
\begin{figure}[ht!]
\centering
\resizebox{0.45\textwidth}{!}{
\includegraphics{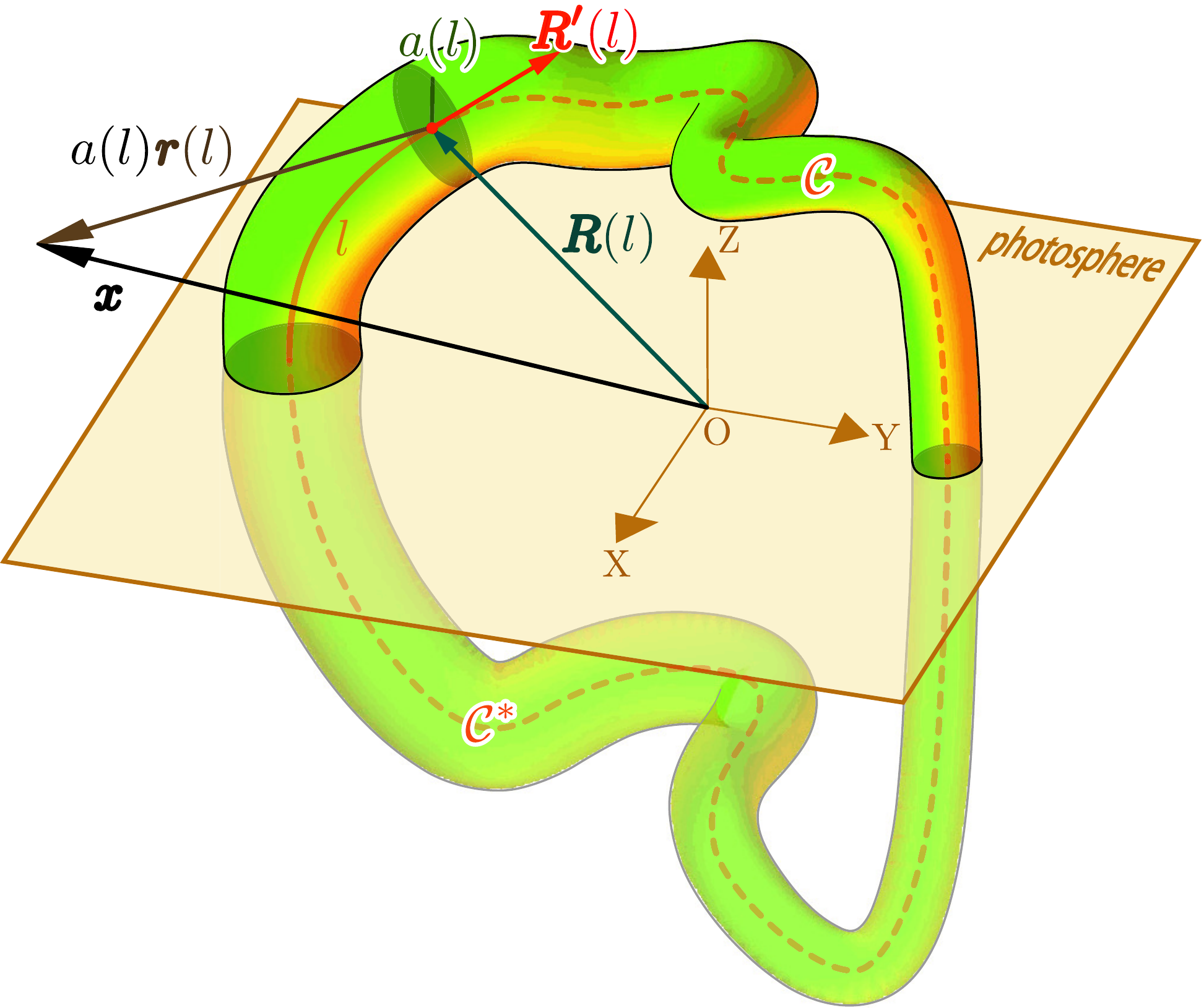}}

\caption{FR with a circular cross-section of radius $a(l)$ and coronal and subphotospheric axis paths $\mathcal C$ and ${\mathcal C}^{*}$, respectively, defined by a radius-vector $ {\pmb R}(l)$, where $l$ is the path arc length.
	\label{f:FR_3d}}
\end{figure}
\begin{eqnarray}
     {\pmb B}_{\mathrm{FR}} &=&  \nabla \times {\pmb A}_{I} +  \nabla \times {\pmb A}_{F} \, , 
	\label{BFR}  \\
     {\pmb A}_{I}({\pmb x}) &=& \frac{\mu I}{4\pi} \int_{\cal C\, \cup\,C^{*}} K_{I}(r) \; {\pmb R}^{\prime}(l) \;  \frac{ {\rm d}l }{  a(l)  }  \, , 
	\label{AI}  \\
     {\pmb A}_{F}({\pmb x}) &=& \frac{F}{4\pi} \int_{\cal C\, \cup\,C^{*}}  K_{F}(r) \; {\pmb R}^{\prime}(l) \times {\pmb r} \;  \frac{ {\rm d}l }{  a(l) ^2 } \, , 
	\label{AF}
\end{eqnarray}
where ${\pmb r} \equiv {\pmb r}(l) = \left( {\pmb x} -{\pmb R}(l) \right) / a(l) $.
The field $ {\pmb A}_{I}({\pmb x})$ and its curl represent the axial vector potential and the azimuthal magnetic field, respectively, generated by a net current $I$.
The field $ {\pmb A}_{F}({\pmb x})$ and its curl represent the azimuthal vector potential and the axial magnetic field, respectively, generated by a net flux $F$.

Both integrals are taken over the coronal path ${\cal C}$ and its subphotospheric counterpart ${\cal C}^{*}$ that closes the current/flux circuit.
The path ${\cal C}^{*}$ can be chosen, for example, as a mirror image of ${\cal C}$ to keep the photospheric normal magnetic field outside the FR unchanged, but other choices of ${\cal C}^{*}$ are possible as well.
Externally, a thin FR manifests itself as a thread carrying an axial current $I$ and axial flux $F$,
which means that Equations (\ref{AI}) and (\ref{AF}) should asymptotically coincide with the classical Biot-Savart laws whose kernels are given by $K_{I}(r) = 1/r$ and $K_{F}(r) = 1/r^{3}$ \citep{Jackson1962}.

We take the latter as exact expressions for our kernels outside the FR and extend them to the FR interior by resolving singularities at $r=0$ and making the interior field approximately force-free.
We do this by simply identifying the kernels of straight and curved force-free FRs of a circular cross-section,
which is allowable if local curvature radii ${\mathcal R}_{\mathrm c}(l)$ of the FR axis are large enough, i.e., ${\mathcal R}_{\mathrm c}(l) \gg a(l) $. 
We call these kernels the regularized Biot-Savart laws (\RBSL) kernels.

In addition, we assume hereafter, for simplicity, that $a(l)\equiv a = \mbox{const}$ along modeled FRs.
This assumption is justified by the successful applications of our \RBSL\ method to configurations with coherent FR structures  (see Section \ref{s:exmpls}).

Let us use $a$ as a length unit for the distances in our consideration,
and let $\rho$ be the distance from $\pmb x$ to the axis of a straight cylindrical FR.
Then, the arc length of the axis is $l=\sqrt{r^2-\rho^2}$.
Changing the integration variable from $l$ to $r$ in Equation (\ref{AI}) and taking the FR cylinder of length $2L$, we obtain
\begin{eqnarray}
  {\pmb A}^{L}_{I}(\rho) = \frac{\mu I \skew{3}\hat{\pmb l}}{4\pi}\; 2 \int_{\rho}^{\sqrt{L^2+\rho^2}} \frac{r\,K_{I}(r)}{\sqrt{r^2-\rho^2}} \; {\mathrm d}r  \, . 
	\label{AIro} 
\end{eqnarray}
As expected, this integral diverges logarithmically in the limit of $L \rightarrow \infty$.
However, it can be ``renormalized'' by subtracting the constant ${\pmb A}^{L}_{I}(1)$, such that it becomes convergent in this limit to yield
\begin{eqnarray}
   2\int_{\rho}^{1} \frac{r\,K_{I}(r)  \; {\mathrm d}r}{\sqrt{r^2-\rho^2}}
 - 2\ln\left(1+\sqrt{1-\rho^2} \right)
 = A_{\mathrm{ax}}(\rho) \, .
          \nonumber \\
	\label{eqKI}
\end{eqnarray}
The integration result is equated here to the axial component, $A_{\mathrm{ax}}(\rho)$, of the vector potential that is normalized by $\mu I/(4\pi)$ \fb{and generally determined up to an arbitrary additive constant}.
\fb{In our $A_{\mathrm{ax}}(\rho)$, however, this constant must be fixed by the condition $A_{\mathrm{ax}}(1)=0$ because of the used ``renormalization''.}

The integral of Equation (\ref{AF}) is convergent for the cylindrical FR and straightforwardly reduces to
\begin{eqnarray}
  2\,\rho \int_{\rho}^{1} \frac{r\,K_{F}(r) \; {\mathrm d}r}{\sqrt{r^2-\rho^2}}
 + \frac{2\,\rho}{1+\sqrt{1-\rho^2}} 
 =  A_{\mathrm{az}}(\rho) \, ,
         \nonumber \\
	\label{eqKF}
\end{eqnarray}
where the integration result is equated to the azimuthal component, $A_{\mathrm{az}}(\rho)$, of the vector potential normalized by $F/(4\pi a)$.
\fb{This normalization automatically implies that $A_{\mathrm{az}}(1)=2$.}

Equations (\ref{eqKI}) and (\ref{eqKF}) are related to the Abel integral equation that can be solved analytically \citep[see p. 531 in][]{Polyanin2008}.
Using this fact, we have found general solutions of these equations in terms of the following quadratures:
\begin{eqnarray}
  K_{I}(r) &=& \frac{2 \arcsin r}{\pi r} 
    - \frac{1}{\pi} \frac{\mathrm d\ }{{\mathrm d}r} \left[ r 
    \int_{r}^{1}  \frac{ A_{\mathrm{ax}}(\rho) \; {\mathrm d}\rho }{ \rho \sqrt{\rho^2 - r^2} } \right] \, , 
            \nonumber \\
	\label{KIgen}\\
  K_{F}(r) &=& \frac{2}{\pi r^{2}} \left(\frac{\arcsin r}{r}-\sqrt{1-r^2}\right)
        \nonumber \\
        &-& \frac{1}{\pi} \frac{\mathrm d\ }{{\mathrm d}r} \left[ r 
    \int_{r}^{1}  \frac{ \left( A_{\mathrm{az}}(\rho) - 2 \rho \right)\; {\mathrm d}\rho }{ \rho^2 \sqrt{\rho^2 - r^2} } \right]\, .
	\label{KFgen}
\end{eqnarray}
They determine the \RBSL\ kernels via given axial and azimuthal components of the vector potential of a cylindrical FR that locally approximates a curved thin FR of any shape.

To make such a curved FR approximately force-free, let us take $A_{\mathrm{ax}}(\rho)$ and $A_{\mathrm{az}}(\rho)$ corresponding to a straight force-free FR.
These components are determined from a force-free equation, for which the profile $j_{\mathrm {ax}}(\rho)$ of the axial current density can be chosen freely.
We take the components derived for a FR with a parabolic $j_{\mathrm{ax}}(\rho)$-profile  ($\rho \in [0,\: 1]$) and vanishing axial field at $\rho = 1$ \citep[cf. Equations (63)--(69) in][]{Titov2014}:
\begin{eqnarray}
  j_{\mathrm{ax}}(\rho) &=& 2\left( 1-\rho^2 \right) \, ,
	\label{jaxpb} \\
  A_{\mathrm{ax}}(\rho) &=& \frac{1}{2} \left( 1 - \rho^2 \right) \left( 3 - \rho^2 \right) \, , 
	\label{Aaxpb}\\
  A_{\mathrm{az}}(\rho) &=& \frac{2\rho}{3\sqrt{3}} \left( 5-2\rho^2 \right)^{3/2} \, ,  
	\label{Aazpb}\\
  F &=& \frac{\pm 3}{5\sqrt{2}} \mu I a,
	\label{Fpb}
\end{eqnarray}
where $j_{\mathrm{ax}}(\rho)$ is normalized by $I/(\pi a^2)$.
Integrating Equations (\ref{KIgen}) and (\ref{KFgen}) for this case, we obtain the corresponding \RBSL\ kernels at $0\le r \le 1$: 
\begin{eqnarray}
K_{I}(r) &=& \frac{2}{\pi} 
     \left( \frac{\arcsin r}{r} + \frac{5-2\, r^2}{3} \sqrt{1-r^2}
     \right) \, , 
	\label{KI}  \\
K_{F}(r) &=& \frac{2}{\pi r^{2}} \left(\frac{\arcsin r}{r}-\sqrt{1-r^2}\right) 
  + \frac{2}{\pi} \sqrt{1-r^2}
  \nonumber \\
  &+&\frac{5-2\, r^2}{2 \sqrt{6}}
      \left[ 1 - \frac{2}{\pi} \arcsin\left( \frac{1+2\,r^2}{5-2\,r^2} \right)
     \right]
      \,.
              \nonumber \\
	\label{KF}
\end{eqnarray}
\begin{figure}[ht!]
\centering
\resizebox{0.45\textwidth}{!}{
\includegraphics{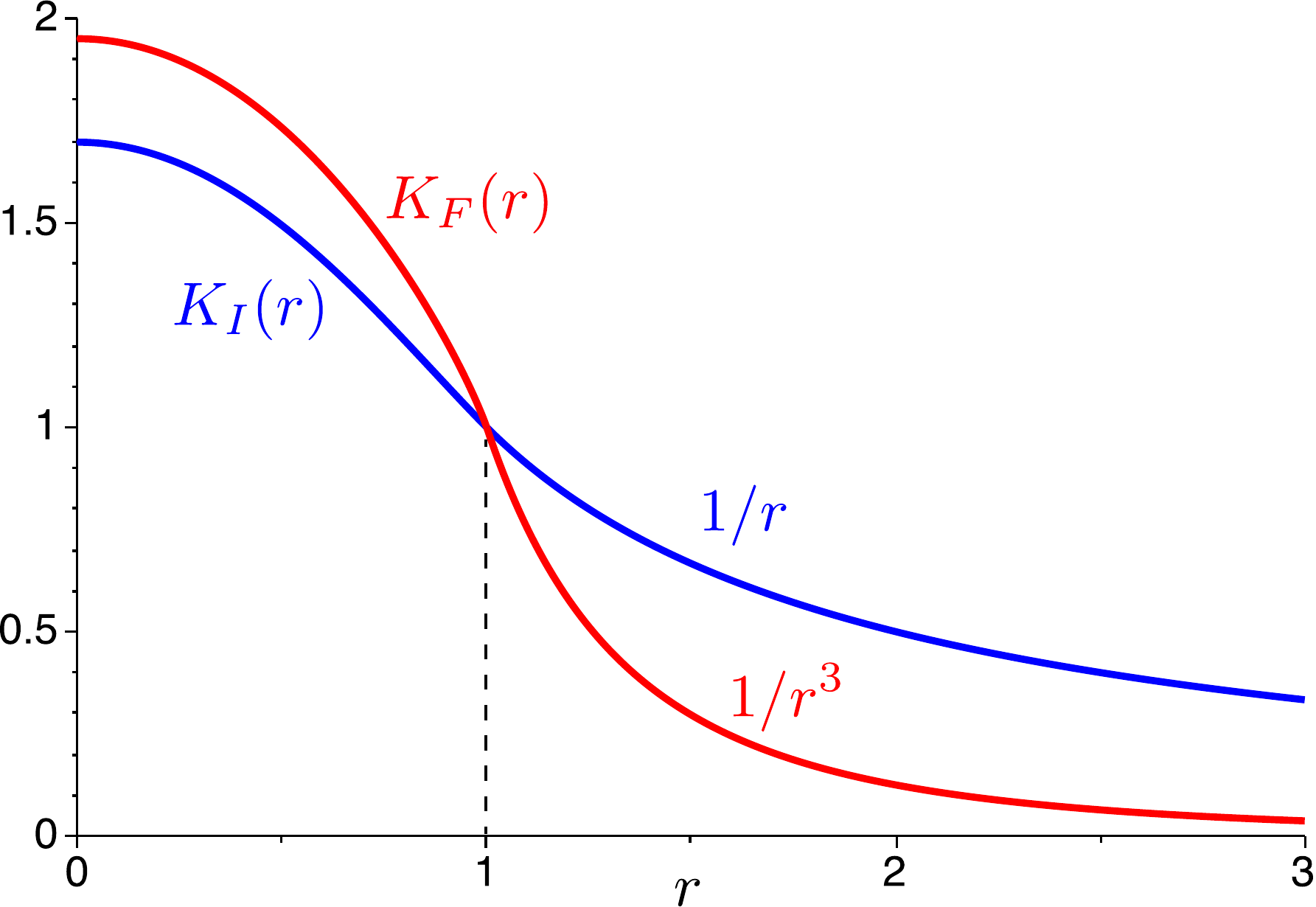}}
\caption{The \RBSL\ kernels for an FR with a parabolic  $j_{\mathrm{ax}}(\rho)$-profile and axial magnetic field vanishing outside the FR;
the inner ($0\le r \le 1$) and outer ($r>1$) solutions smoothly conjugate to each other.
	\label{f:KIKF}}
\end{figure}
Figure \ref{f:KIKF} shows that they are smoothly conjugate to the classical Biot-Savart kernels outside the FR, as required.

Equations (\ref{AI}) and (\ref{AF})  together with the found \RBSL\ kernels reduce the computation of  $ {\pmb A}_{\mathrm{FR}} \equiv \ {\pmb A}_{I} + {\pmb A}_{F} $ to the calculation of two line integrals.
If one computes $ {\pmb A}_{\mathrm{FR}}  $ on a numerical grid,
the key advantage of our method is that the \RBSL\ integration path is the same for all grid points.


\section{Illustrative Examples}
	\label{s:exmpls}

We have implemented our method for magnetic configurations defined on numerical grids under different assumptions on  closing the axis path $\mathcal C$ by $\mathcal C^{*}$.
\fb{Below we present several tests of the method to verify its capacity to construct approximately force-free FR configurations.
To assess how close these configurations are to equilibrium, we relax them by using our spherical MHD code, MAS \citep{Lionello2009}, in zero-$\beta$ mode \citep{Mikic2013a}, and then visually compare the initial and final field-line structures. 
}

\subsection{Test Case 1: TDm Model} 
	\label{s:t1}
Our first test is intended to reproduce a 
TDm configuration that includes a toroidal FR with a parabolic $j_{\rm ax}(\rho)$-profile given by Equation (\ref{jaxpb}).
\begin{figure}[ht!]
\centering
\resizebox{0.45\textwidth}{!}{
\includegraphics{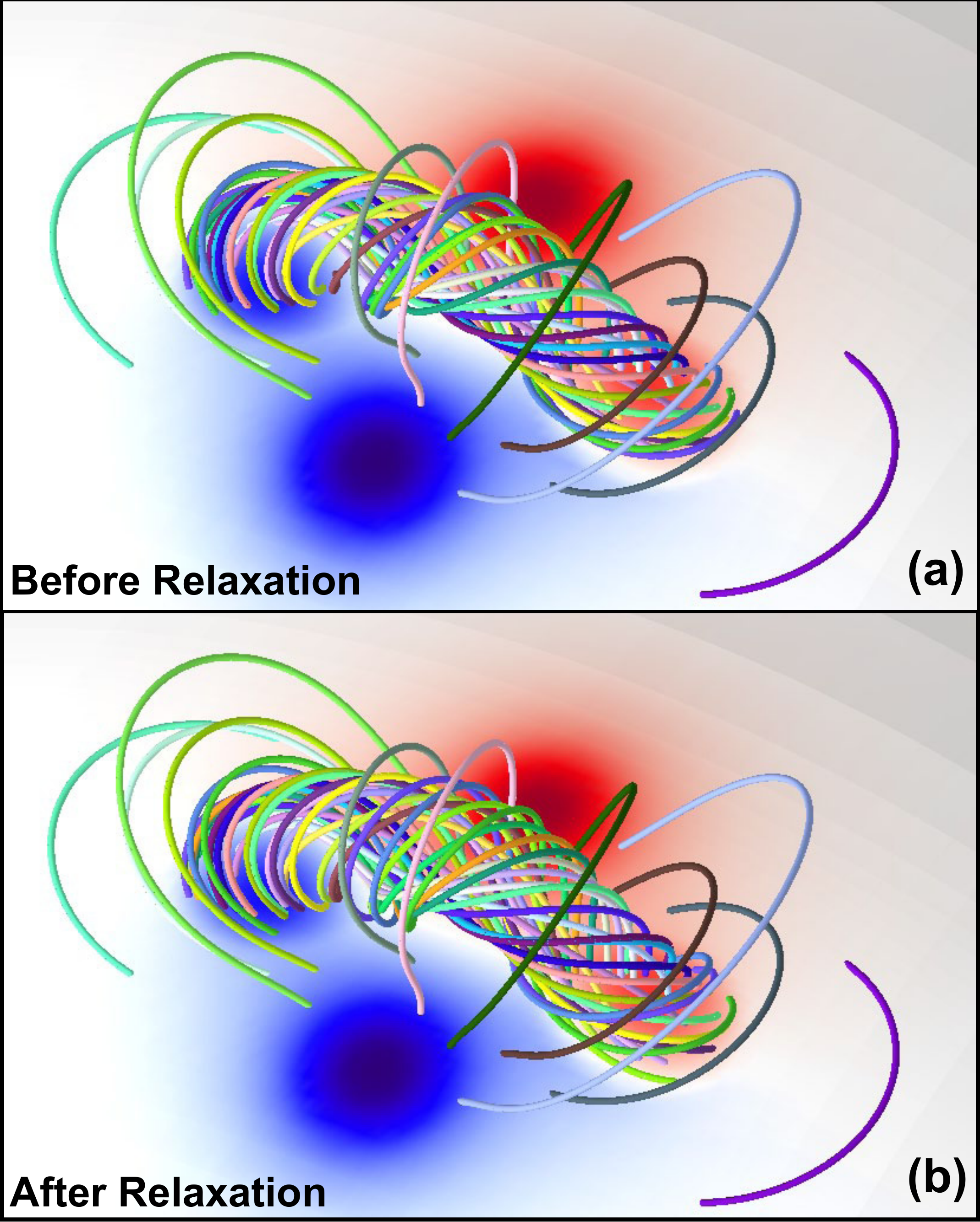}}
\caption{Reproducing the TDm model (test case 1): the \RBSL\ configuration before (a) and after (b) zero-$\beta$  MHD relaxation.
The field lines in both configurations have the same footpoints and colors; the boundary radial field $B_{r0}$ is colored in blue ($B_{r0}<0$) and red ($B_{r0}>0$).
	\label{f:TDm}}
\end{figure}
The FR is embedded in an idealized bipolar background field ${\pmb B}_{q}$, which is modeled by two fictitious point sources of strength $|\pm q|=100\:{\mbox{T\:Mm}^{2}}$ placed at a depth of $50\: \mbox{Mm}$ below the photospheric boundary and at a distance of $150\:\mbox{Mm}$ from each other.
The axis path $\mathcal C$ follows an iso-contour, $B_{q} = {\mathrm{const}}$, of a circular-arc shape in the vertical plane of symmetry of the configuration, and is closed by a subphotospheric arc $\mathcal C^{*}$ to form a circle of radius $\mathcal R_{\mathrm c} = 110\: \mbox{Mm}$.
The torus minor radius is set to $a=45\:\mbox{Mm}$, which together with the chosen iso-contour determines the parameters $I$ \citep[Equation (7) in][]{Titov2014} and $F$ (Equation (\ref{Fpb})).

We first compared the initial \RBSL\ configuration to the equivalent TDm version using the same fluxes, geometry, and parabolic current profile. 
We found that their vector potentials differ by less than 2\%,
indicating that this choice for the \RBSL\ kernels indeed matches the TDm formulation for circular FRs.  

We then compare the initial FR configuration (Figure \ref{f:TDm}(a)) to one produced after a line-tied zero-$\beta$ MHD relaxation (Figure \ref{f:TDm}(b)).
In spite of a relatively large FR curvature, $a/{\mathcal R_{\mathrm c}}= 0.41$, the relaxation yields a numerical force-free field that is almost identical to the initial \RBSL\ configuration, demonstrating that the force-freeness property extends nicely from straight to curved FRs.

\subsection{Test Case 2: Sigmoidal Configuration of  the 2009 February 13 CME Event} 
	\label{s:t2}

Our second test is designed to benchmark the method for simple yet realistic magnetic configurations.
For this purpose, we choose the 2009 February 13 CME event where the pre-eruptive magnetic field
had a characteristic sigmoidal structure above the polarity inversion line (PIL) of the source region \citep{Miklenic2011}.
We do not intend to perfectly reproduce this structure, or preserve the radial component of the photospheric field, $B_{r0}$, obtained from observations. 
Rather, our purpose is to check whether the \RBSL\ method can produce a similar sigmoidal structure by simply superimposing ${\pmb B}_{\mathrm{FR}}$ and the potential field ${\pmb B}_{\mathrm{P}}$ extrapolated from the $B_{r0}$-map. 
\begin{figure*}[ht!]
\centering
\resizebox{0.95\textwidth}{!}{
\includegraphics{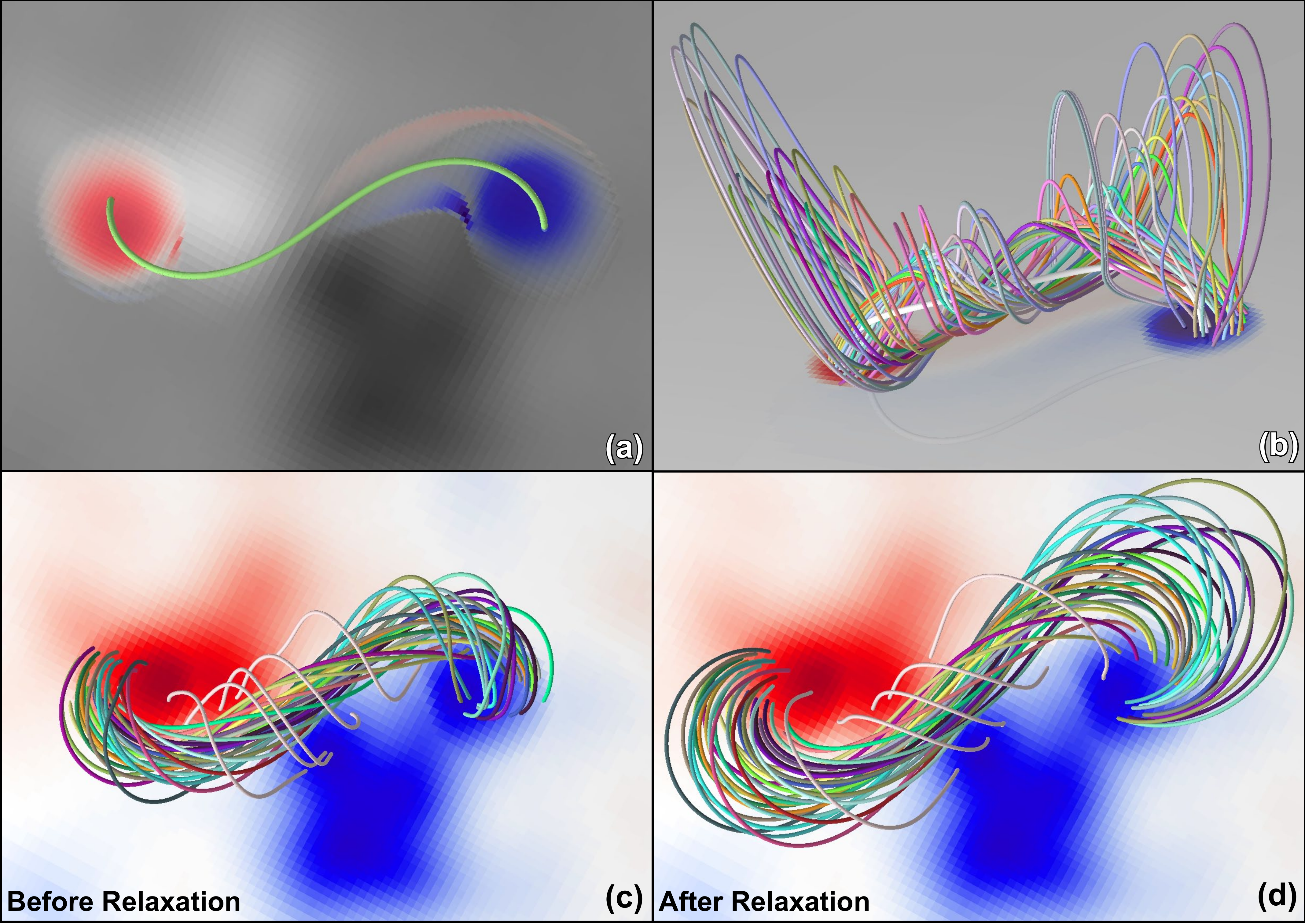}}
\caption{Steps of modeling the pre-eruptive configuration of the 2009 February 13 CME event (test case 2): (a) top view of the chosen FR-axis path (green line) in the corona with the grayscaled photospheric $B_{r0}$-map (full magnetic field; black: $B_{r0}<0$; white: $B_{r0}>0$) and the $j_{r0}$-map in  blue ($j_{r0}<0$) and red ($j_{r0}>0$);  (b) side view of the FR-axis path (thick white line) and field lines for the FR-field only, with the corresponding semi-transparent blue-red $B_{r0}$-map.
The \RBSL\ configuration before (c) and after (d) zero-$\beta$ MHD relaxation; the $B_{r0}$-map is colored in blue and red.
	\label{f:2009Feb13_1}}
\end{figure*}

We first choose an S-shaped axis path $\mathcal{C}$ above the PIL of the $B_{r0}$-map (Figure \ref{f:2009Feb13_1}(a)), as suggested by the observations.
This path is closed by a subphotospheric path $\mathcal{C}^{*}$
that mirrors the path $\mathcal{C}$ about the local horizontal plane passing through the footpoints of $\mathcal{C}$. The superposition of ${\pmb B}_{\mathrm{FR}}$ and ${\pmb B}_{\mathrm{P}}$ will naturally modify the $B_{r0}$-map inside the FR footprints due to the axial flux (Figure \ref{f:2009Feb13_1}(b)), but such a path mirroring causes most of the radial component of the azimuthal FR field to vanish at the photosphere. Stripes of weak $B_{r0}$ remain because of the small spherical curvature of the boundary, but this could, in principle, be eliminated by a small adjustment of $\mathcal{C}^{*}$.

The equilibrium axial current $I$ is estimated in two steps.
First, for some current $I_{0}$ and a middle point of the axis-path ${\pmb R}^{*}$, we calculate the potential field
${\pmb B}^{*}_{\mathrm{P}} \equiv {\pmb B}_{\mathrm{P}}({\pmb R}^{*})$ 
and the azimuthal field ${\pmb B}^{*}_{I_{0}} \equiv  {\pmb B}_{I_{0}}({\pmb R}^{*}) \equiv \left . \nabla \times {\pmb A}_{I_{0}} \right|_{{\pmb x}= {\pmb R}^{*} }$.
Since $|{\pmb B}^{*}_{\mathrm{P}}  + c \, {\pmb B}^{*}_{I_{0}}|$ as a function of $c$ has a minimum value at $c = c_{0} \equiv - \left( {\pmb B}^{*}_{\mathrm{P}} {\pmb \cdot} {\pmb B}^{*}_{I_{0}} \right) / { B}^{*2}_{I_{0}}$, we obtain the desired estimate for the equilibrium current $I = c_{0}\, I_{0}$.

Figure \ref{f:2009Feb13_1}(c) shows that
the resulting field ${\pmb B}_{\mathrm{FR}} + {\pmb B}_{\mathrm{P}}$
indeed contains a sigmoidal FR.
The sigmoid expands during line-tied MHD relaxation to form a stable FR of a more pronounced S-shape (Figure  \ref{f:2009Feb13_1}(d)).

By changing the coefficient $c$ in the linear superposition $c {\pmb B}_{\mathrm{FR}} + {\pmb B}_{\mathrm{P}}$, one can easily generate a family of solutions with sigmoidal FRs that carry different axial current $I$ and flux $F$.
This is very important for parameter studies, in which, e.g., the critical parameters for the onset of eruption are needed to be found.  

\subsection{Test Case 3: Pre-eruptive Configuration of the 2011 October 1 CME Event} 
	\label{s:t3}

\begin{figure}[ht!]
\centering
\resizebox{0.45\textwidth}{!}{
\includegraphics{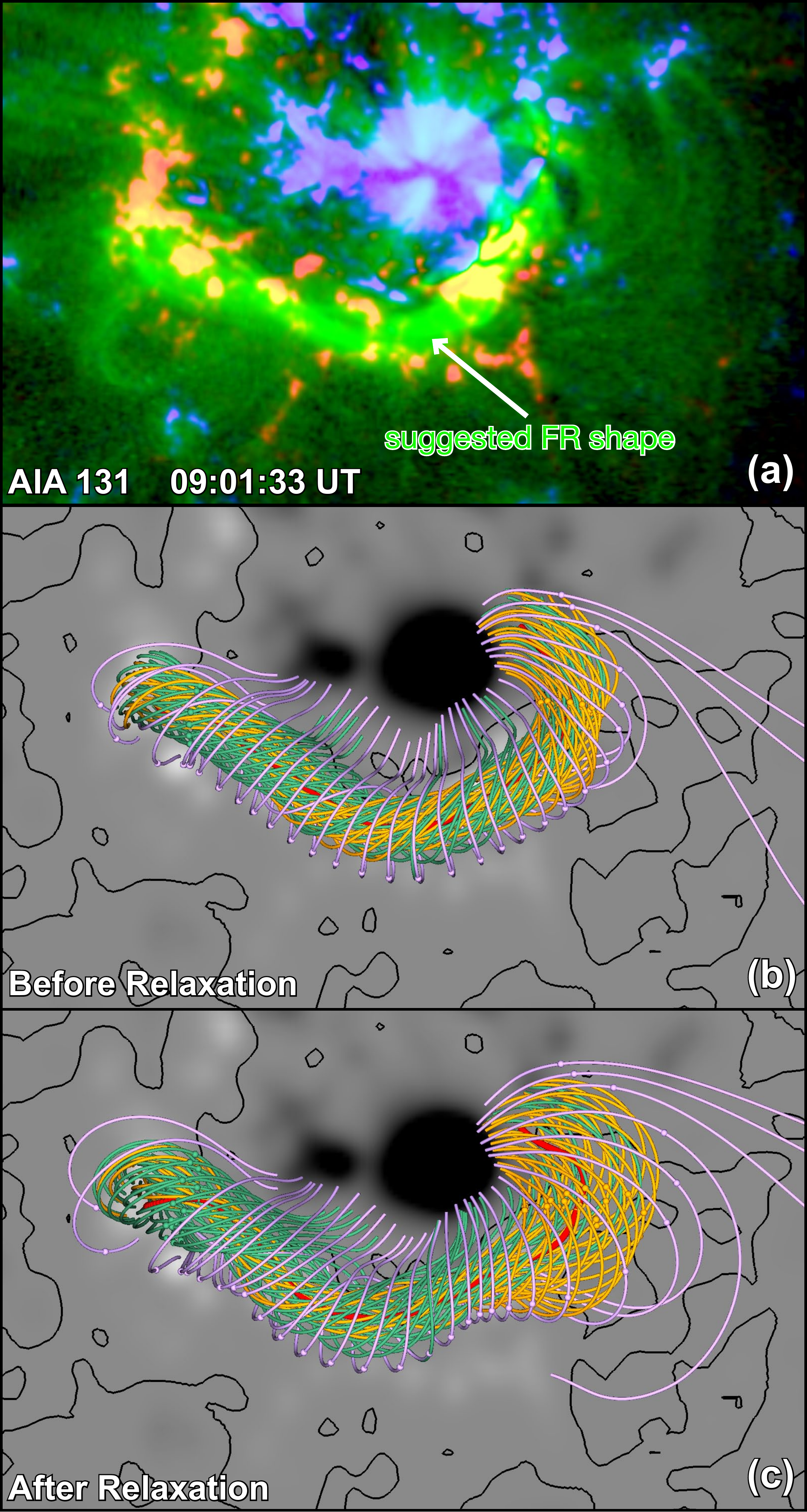}}
\caption{Modeling the pre-eruptive configuration of the 2011 October 1 CME event (test case 3): (a) FR-like structure suggested by the AIA 131\:\AA\ image superimposed on the photospheric blue-red $B_{r0}$-map; the \RBSL\  configuration before (b) and after (c) zero-$\beta$ MHD relaxation; orange and green field lines start near the FR footprints, purple field lines show a magnetic arcade enclosing the FR, whose axis is shown by a red line; field lines are drawn by tracking the motion of selected fluid elements (balls) in time; the same elements are used in (b) and (c).
	\label{f:2011Oct1}}
\end{figure}

Our third test explores how well the method works for more complex configurations, such as the one that produced the 2011 October 1 CME \citep[Figure \ref{f:2011Oct1}(a);][]{Temmer2017}. 
This configuration had a PIL that separated a strong (negative) sunspot and a weak dispersed (positive) flux concentration. The strong inhomogeneity of the ambient magnetic field near the PIL poses a serious challenge to embedding a force-free FR into this region.

We first construct the FR-axis path by using SDO/AIA 131\:\AA\  observations, which show a bright, curved, elongated feature over the PIL prior to the eruption (Figure \ref{f:2011Oct1}(a)). This projection does not constrain the height, so we chose a height that was slightly larger than the suggested width of the feature. To improve the match between the observed structure and the modeled pre-eruptive FR, we also explored different types of closures for the FR-axis path.

Figure \ref{f:2011Oct1}(b) presents our best solution. In contrast to test case 2, the observed $B_{r0}$-distribution is preserved using a technique similar to \citet{vanBall2004}. We do this by removing the non-vanishing radial component of the \RBSL\ field at the photosphere from the original $B_{r0}$-distribution and calculating the corresponding potential field.
Superimposing this and FR fields ensures that the radial field at the boundary exactly matches $B_{r0}$. 

The coronal axis path $\mathcal C$ is fine-tuned to minimize a residual Lorentz force along the embedded FR,
by repeating small perpendicular displacements of $\mathcal C$ in the  directions that yield the strongest decrease of this force.
Figures \ref{f:2011Oct1}(b) and \ref{f:2011Oct1}(c) show that the configuration with the fine-tuned FR is similar to the force-free equilibrium reached in the line-tied MHD relaxation.


\section{Summary}
	\label{sum}

We have developed a new method for constructing force-free FRs embedded into potential magnetic fields.
Our method allows one to use an arbitrary FR axis shape and to estimate the equilibrium parameters from the background field, making it generally applicable and computationally efficient.

The FR field is expressed in terms of the axial and azimuthal vector potentials defined by the \RBSL s for a given FR axis, total axial current and axial flux.
The axis shape is determined by following the PIL of an eruption's source region, using observed magnetograms, and by using observations of this region.
The height variation along the axis and other FR parameters are estimated via potential field extrapolation.
The FR-axis shape can be iteratively adjusted to minimize the Lorentz force along the FR, after which the configuration is subjected to line-tied MHD relaxation toward a numerical equilibrium.

We successfully tested our method for the TDm model \citep{Titov2014} and the pre-eruption configurations of the 2009 February 13 and 2011 October 1 CME events.
Our tests demonstrate that the \RBSL\ method is a very flexible and efficient way to construct coherent flux-rope structures of non-trivial geometry.
We envision that this method will be particularly useful for theoretical studies of FRs with complex geometries, and for initializing data-constrained simulations of solar flares and CMEs.
We are currently extending this method for modeling FRs with variable cross-sections, which will further increase its flexibility and allow one to initialize interplanetary CME simulations as well.


\acknowledgments

This research was supported by NASA's HSR, LWS, and HGI programs,
NSF grants AGS-1560411 and AGS-1135432,
and AFOSR contract FA9550-15-C-0001.
Computational resources were provided by NSF's XSEDE and NASA's NAS.


\bibliographystyle{apj}

\end{document}